\documentclass[aps,amssymb,amsmath,
twocolumn,showpacs,superscriptaddress,
groupedaddress]{revtex4-1}

\usepackage{tikz}
\usepackage{natbib}
\usetikzlibrary{positioning}
\usepackage{graphicx}
\usepackage{caption}
\usepackage{subcaption}
\usepackage[section]{placeins}
\usepackage{subfig}
\usepackage{dcolumn}
\usepackage{bm}   
\definecolor{OliveGreen}{cmyk}{0.64,0,0.95,0.40}
\definecolor{royalfuchsia}{rgb}{0.79, 0.17, 0.57}
\definecolor{vividauburn}{rgb}{0.58, 0.15, 0.14} 

\begin{document}
\title{Benchmark values for molecular three-center integrals arising in the Dirac equation}
\author{A. Ba{\u g}c{\i}}
\email{albagci@univ-bpclermont.fr}
\author{P. E. Hoggan}
\affiliation{Institute Pascal, UMR 6602 CNRS, University Blaise Pascal, 24 avenue des Landais BP 80026, 63177 Aubiere Cedex, France}

\begin{abstract}
The authors in their previous papers obtained compact, arbitrarily accurate expressions for two-center one- and two-electron relativistic molecular integrals expressed over Slater-type orbitals. In this present study, the accuracy limits of given expressions is examined for three-center nuclear attraction integrals, which are the first integral set do not have analytically closed form relations. They are expressed through new molecular auxiliary functions obtained via Neumann expansion of Coulomb interaction. The numerical global adaptive method is used to evaluate these integrals for arbitrarily values of orbital parameters, quantum numbers. Several methods, such as Laplace expansion of Coulomb interaction, single-center expansion, Fourier transformation method, have been performed in order to evaluate these integrals considering the values of principal quantum numbers in the set of positive integer numbers. This is the first attempts to study the three-center integrals without any restrictions on quantum numbers and in all ranges of orbital parameters.
\begin{description}
\item[Keywords]

\item[PACS numbers]
... . 
\end{description}
\end{abstract}
\maketitle

\section{\label{sec:intro}Introduction}
The LCAO-SCF \cite{Roothaan1951} method is generally employed for molecules, in which molecular wave functions taken to be linear combinations of atomic basis functions whose should possess the cusps condition at the nuclei \cite{Kato1957} and decay exponentially for large distances \cite{Agmon1982}. This approach leads to use, namely, Slater-type orbitals \cite{Slater1930, Parr1957},
\begin{align} \label{eq:STSOs}
\chi_{nlm} \left(\zeta,\vec{r}\right)=
\frac{\left(2\zeta \right)^{n+1/2}}{\sqrt{\Gamma(2n+1)}}r^{n-1}e^{-\zeta r}Y_{lm}(\theta,\phi),
\end{align}
here, $Y_{lm}$ are complex or real spherical harmonics $(Y^{*}_{lm}=Y_{l-m}; Y_{lm} \equiv S_{lm})$ differs from the Condon$-$Shortley phases by sign factor $(-1)^{m}$ \cite{CS1935, Steinborn1978, Blanco1997}, $\Gamma(z)$ are gamma functions \cite{Abramowitz1972}, $\left\lbrace n, l, m \right\rbrace$ are the principal, orbital, magnetic quantum numbers with, $n \in \mathbb{R}^{+}$, $0\leq l \leq \lfloor n \rfloor$, $-l \leq m \leq l$ and $\lfloor n \rfloor$ stands for the integer part of $n$, respectively, in one$-$ and two$-$electron multi$-$center molecular integrals. These integrals needs to be calculated in spectroscopic accuracy in order to meaningful discussions on basis-set expansion methods, Born-Oppenheimer energy, vibrational frequency calculations. The difficulty of finding analytically closed form relations, however, for molecular integrals have more than two-center referred to as \textit{The bottleneck of quantum chemistry} \cite{Mulliken1959}, have been greatest obstacle since Slater-type orbitals have no simple addition theorem; relations for products of two Slater-type orbitals centered on different positions not available in compact form \cite{Bouferguene1998}.

The Slater-type orbitals are obtained by simplification of Laguerre functions in hydrogen$-$like orbitals \cite{Willock2009} by keeping only the term of the highest power of $r$, for integer values of principal quantum number $n$ (ISTOs), where $n \in \mathbb{N}^{+}$, $\Gamma(2n+1)=(2n)!$ and it has been proved that they provide extra 
flexibility for closer variational description of atoms and molecules by considering the values of $n$ in more general set of number, namely positive real numbers (NSTOs), where $n \in \mathbb{R}^{+}$. The studies on the evaluation of molecular integrals, thus, are performed in two main group: those restrict the principal quantum number with integer values, which are practically used in nonrelativistic molecular electronic structure calculations \cite{Bouferguene1996, Rico2001} and those free them from any specification but also reduce the area of applications only to investigation of atoms \cite{Koga1997-1, Koga1997-2, Koga1997-3, Koga1998, Koga2000, Erturk2015}.

The multi$-$center molecular integrals over ISTOs can be evaluated by expansion of Slater-type orbitals through complete orthonormal basis functions to a new origin \cite{Barnett1951, Harris1965, Guseinov1978, Guseinov2001, Bouferguene2005} (see also references therein),
\begin{multline} \label{eq:WAVEEXPTHEO}
\chi_{nlm}(\zeta,\vec r_{A})\\
=\lim_{N_{e} \to \infty}\sum_{n'l'm'}^{N_{e}} V_{nlm,n'l'm'}^{N_{e}}(\zeta,\vec R_{AB})\chi_{n'l'm'}(\zeta,\vec r_{B}).
\end{multline}
or by expressing them as a finite linear combination of $B$ functions through Fourier transform \cite{Filter1978-1, Filter1978-2, Weniger1983, Grotendorst1985, Steinborn1992, Homeier1992}. However, infinite series representation formulas arising in expansion method require increasing upper limit of summation as much as possible to converge to exact values with sufficient decimals (the choice adopted as threshold for the total energy in nonrelativistic variational energy calculation is of order E$-$03 atomic units, therefore, constitute matrix elements should be accurate to E$-$10 atomic units) and presence of spherical Bessel functions brings computational difficulties in Fourier transform method since they provoke an oscillation \cite{Safouhi1999, Safouhi2000, Safouhi2003-1, Safouhi2003-2}.  
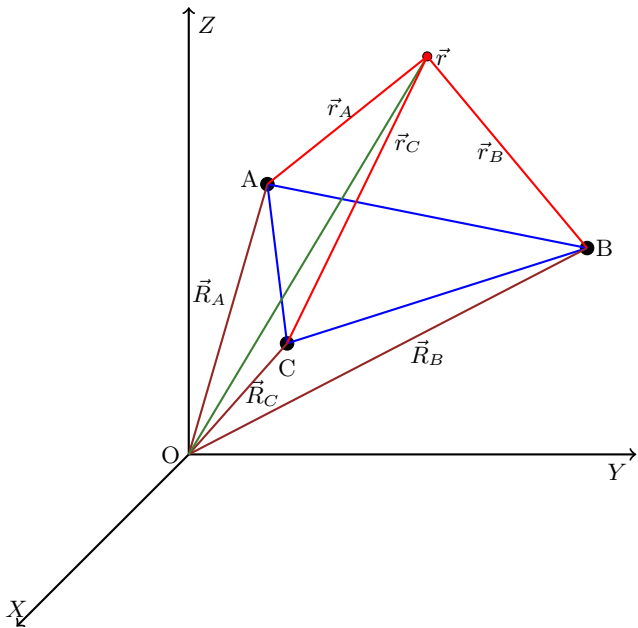
\begin{figure}[htp!]
\centering
\begin{tikzpicture}[scale=0.85]
\coordinate (Origin) at (0,0,0);
\draw[thick,->] (0,0,0) -- (7,0,0) node[anchor=north east]{$Y$};
\draw[thick,->] (0,0,0) -- (0,7,0) node[anchor=north west]{$Z$};
\draw[thick,->] (0,0,0) -- (0,0,7) node[anchor=south]{$X$};

\coordinate (A) at (2,5,2);
\coordinate (B) at (7,4,2);
\coordinate (C) at (2.5,2.7,2.5);
\coordinate (E) at (4.5,7,2);

\draw [fill=blue] (Origin) circle  node [left] {O};
\draw [fill=black] (A) circle (3pt) node [below left=-0.3cm and 0.00 of A] {A};
\draw [fill=black] (B) circle (3pt) node [right] {B};
\draw [fill=black] (C) circle (3pt) node [below=0.1cm and 0.00 of C] {C};
\draw [fill=red] (E) circle (2pt) node [right] {$\vec{r}$};

\draw[blue,thick] (2,5,2) -- (7,4,2);
\draw[blue,thick] (2,5,2) -- (2.5,2.7,2.5);
\draw[blue,thick] (7,4,2) -- (2.5,2.7,2.5);

\draw[vividauburn,thick] (0,0,0) -- (2,5,2) node [black, pos=0.6, left] {$\vec{R}_{A}$};
\draw[vividauburn,thick] (0,0,0) -- (7,4,2)node [black,pos=0.6, below] {$\vec{R}_{B}$};
\draw[vividauburn,thick] (0,0,0) -- (2.5,2.7,2.5) node [black,pos=0.75, below] {$\vec{R}_{C}$};
\draw[OliveGreen,thick] (0,0,0) -- (4.5,7,2) node [black,pos=0.55, left] {};

\draw[red,thick] (2,5,2) -- (4.5,7,2) node [black,pos=0.6, left] {$\vec{r}_{A}$};
\draw[red,thick] (7,4,2) -- (4.5,7,2) node [black,pos=0.6, below] {$\vec{r}_{B}$};
\draw[red,thick] (2.5,2.7,2.5) -- (4.5,7,2) node [black,pos=0.7, right] {$\vec{r}_{C}$};

\coordinate (OriginA) at (2,5,2);

\coordinate (OriginB) at (7,4,2);

\coordinate (OriginC) at (2.5,2.7,2.5);
\end{tikzpicture}
\caption {\label{fig:GeometryFig} Depiction of the coordinates for motion of an electron in the field of three stationary Coulomb centers, namely $A$, $B$, $C$, where $A=\left\lbrace Z_{A}, Y_{A},X_{A} \right\rbrace$, $B=\left\lbrace Z_{B},Y_{B},X_{B} \right\rbrace$, $C=\left\lbrace Z_{C},Y_{C},X_{C} \right\rbrace$, $\left\lbrace Z, Y, X \right\rbrace$ are the axes of Cartesian coordinates.}
\end{figure}

The problem of multi$-$center integrals evaluation by the use of NSTOs even much more through insurmountable. The Slater type orbitals with noninteger principal quantum numbers do not have infinite series representation formulas; they can not be expanded via complete orthonormal basis functions since power series for a function such as $z^\rho$, $z \in \mathbb{C}$ and $\rho \in \mathbb{R}/\mathbb{N}_{0}$ are not analytic at the origin \cite{Weniger2008, Weniger2012}, where the symbols $\mathbb{C}$, $\mathbb{R}$, $\mathbb{N}_{0}$ used to denote the sets of complex, real and natural numbers, respectively. It should be noted that, this also eliminates possibility of applying binomial expansion theorem in order to evaluate the two-center integrals, those are analytically closed form relations may obtain. Therefore, in mathematical point of view evaluation of multi$-$center molecular integrals using noninteger principal quantum numbers in Slater-type orbitals is an open question. It is far more than better representation of electronic wave$-$function in nonrelativistic electronic structure calculations it is also directly related with solution of the Dirac equation in algebraic approximation. The basis functions to be used in solution of matrix form of the Dirac equation are obtained analogously to L-spinors \cite{Grant2000, Grant2007} which are related to the Dirac hydrogenic solutions. Their explicit form include power functions $r^{\gamma}$, 
\begin{align}
\gamma=\sqrt{\kappa^2-\frac{Z^2}{c^2}}, 
\end{align}
with, $Z$ is nuclear charge, $c$ is speed of light, $\kappa=\pm1,\pm2, \pm2, ...$, respectively. They can only be represent by finite summation of Slater-type orbitals with noninteger principal quantum numbers. 

In particular, the three$-$center integrals are the first set of multi$-$center integrals do not have analytically closed form relations. They have a fundamental importance in the study of molecular systems through $ab-$initio and density functional theory. They are central to the understanding of multi-center integrals. They have been commonly studied with methods presented above. They can also be evaluated through Neumann \cite{Guseinov1976, Rico1992, Harris2002, Peuker2008} and Laplace expansion \cite{Roberts1969, Rico1989, Rico1991} of Coulomb interaction in prolate spheroidal coordinates.

In a new approach the two$-$center integrals have been calculated by the authors for arbitrary values of parameters and quantum numbers via numerical integration techniques \cite{Bagci2014, Bagci2015}. The new relativistic molecular auxiliary functions in prolate spheroidal coordinates are presented \cite{Bagci2015}. They are used to obtain compact form relations for two-electron integrals. Afterwards this idea adapted to calculate overlap integrals via Fourier transform formulas \cite{Silverstone2014}. The same accuracy, 36-digits, is achieved in both methods. These are so far only known precise calculations for molecular integrals over NSTOs. Hence, they are used in this paper to produce benchmark values for three-center one-electron molecular Dirac integrals as a first time in the literature. The {\sl Mathematica} programming language \cite{Mathematica} is utilized for both analytical and numerical calculations.

\section{\label{sec:Threecenter}Three-center nuclear attraction integrals}
Taking into account Fig. \ref{fig:GeometryFig}, where depiction of coordinates are given for one electron in a triangular conformation, the three-center nuclear attraction integrals are defined as follows,
\begin{multline} \label{eq:THREECENTERIDEF}
I_{nlm,n'l'm'}(\zeta,\zeta',\vec{R}_{AB},\vec{R}_{AC})\\
=\int \chi_{nlm}^{*} \left(\zeta,\vec{r}\right) \frac{1}{\vert \vec{r}-\vec{R}_{AC} \vert} \chi_{n'l'm'} \left(\zeta',\vec{r}-\vec{R}_{AB}\right)dV,
\end{multline}
with, $A,B,C$ are three arbitrary points of the euclidian space, $\vec{R}_{AB}=\vec{AB}$, $\vec{R}_{AC}=\vec{AC}$. 

The Neumann expansion for $1/{\vert \vec{r}-\vec{R}_{AC} \vert}$ in prolate spheroidal coordinates ($\xi, \nu, \phi$), where $1\leq\xi\leq\infty$, $-1\leq\nu\leq1$, $0\leq\phi\leq2\pi$ \cite{Ruedenberg1951},
\begin{multline} \label{eq:NEUMANNEX}
\frac{1}{\vert \vec{r}-\vec{R}_{AC} \vert}
=\frac{8\pi}{R_{AB}}\sum_{LM}(-1)^{M}\frac{(L-\vert M \vert)!}{(L+\vert M \vert)!}\\
\times\mathcal{P}_{L}^{\vert M \vert}(\xi_{<})\mathcal{Q}_{L}^{\vert M \vert}(\xi_{>})\\
\times Y_{L}^{M}(\nu_{C},\phi_{C})Y_{L}^{M}(\nu,\phi)^{*},
\end{multline}
here, $\mathcal{P}_{L}^{\vert M \vert}(\xi)$, $\mathcal{Q}_{L}^{\vert M \vert}(\xi)$ are first and second kind associated Legendre functions \cite{Abramowitz1972}, $\left\lbrace \xi_<, \xi_> \right\rbrace$ refers to lesser and greater of $\left\lbrace \xi_c, \xi \right\rbrace$, respectively, is utilized in order to obtain expressions for the three-center nuclear attraction integrals over Slater-type orbitals, where the principal quantum numbers are free from specifications;
\begin{multline}\label{eq:METHREECENTER}
I_{nlm,n'l'm'}(\zeta,\zeta',\vec{R}_{AB},\vec{R}_{AC})\\
=\frac{4\sqrt{2\pi}}{R_{AB}}\mathcal{N}_{nn'}(\zeta,\zeta',R_{AB})\sum_{LM}(-1)^{M} \frac{(L-\vert M \vert)!}{(L+\vert M \vert)!} A_{mm'}^{M}\\
\times Y_{L}^{M}(\nu_{C},\phi_{C}) \left\lbrace \mathcal{Q}_{L}^{\vert M \vert}(\xi_{C}) \mathcal{J}_{nlm,n'l'm'}^{LM}(\zeta,\zeta',R_{AB},\xi_{C}) \right. \\
+ \left. \mathcal{P}_{L}^{\vert M \vert}(\xi_{C}) \mathcal{K}_{nlm,n'l'm'}^{LM}(\zeta,\zeta',R_{AB},\xi_{C})\right\rbrace,
\end{multline}
here,
\begin{multline}\label{eq:NORMCOEF}
\mathcal{N}_{nn'}(\zeta,\zeta',R)=\\
\frac{\left(2\zeta \right)^{n+1/2}\left(2\zeta' \right)^{n'+1/2}}{\left[\Gamma(2n+1)\Gamma(2n'+1) \right]^{1/2}}\left( \frac{R}{2}\right)^{n+n'+1}
\end{multline}
are the normalization constants and $A^{M}$ coefficients \cite{Guseinov1970}
\begin{multline}\label{eq:NORMCOEF}
A_{mm'}^{M}=\frac{1}{\sqrt{2}}\left(2-\vert \eta_{mm'}^{m-m'} \vert \right)^{1/2}\delta_{M,\epsilon\vert m-m' \vert}\\
+\frac{1}{\sqrt{2}}\eta_{mm'}^{m+m'}\delta_{M,\epsilon\vert m+m' \vert},
\end{multline}
are the integration over azimuthal angle. The symbol $\epsilon$ may have the value $\pm 1$ and is determined by the product of the signs $m$ and $m'$ (the sign of zero is regarded as positive). The symbols $\eta_{mm'}^{m\pm m'}$ may have the values $\pm 1$ and 0: if among the indices $m$, $m'$ and $m \pm m'$ there occurs a value equal to zero, then $\eta_{mm'}^{m\pm m'}$ is also zero; if all the indices differ from zero, $\eta_{mm'}^{m\pm m'}= \pm 1$ and the sign is determined by product of the signs $m$, $m'$, $m \pm m'$. Thus, the coefficients $A^{M}$ differ from zero only with the values $\vert M \vert =\vert m-m' \vert$, $\vert M \vert =\vert m+m' \vert$.\\
Since it is assumed axes of prolate spheroidal coordinate system centered on A, B substitutions in Eqs. (\ref{eq:THREECENTERIDEF}, \ref{eq:NEUMANNEX}) can be written as follows \cite{Gordadse1935},
\begin{align}
\xi=\frac{r_{A}+r_{B}}{R_{AB}}; \hspace{5mm}\nu=\frac{r_{A}-r_{B}}{R_{AB}},
\end{align}
\begin{align}
r_{A}=\frac{R_{AB}}{2}\left(\xi +\nu \right); \hspace{5mm}r_{B}=\frac{R_{AB}}{2}\left(\xi -\nu \right),
\end{align}
\begin{align}
\xi_{C}=\frac{R_{AC}+R_{BC}}{R_{AB}}; \hspace{5mm}\nu_{C}=\frac{R_{AC}-R_{BC}}{R_{AB}}.
\end{align}
The $\mathcal{J}^{LM}, \mathcal{K}^{LM}$ integrals are the auxiliary functions and they are defined as,
\begin{multline}\label{eq:JGENAUXILIARY}
\mathcal{J}_{nlm,n'l'm'}^{LM}(\zeta,\zeta',R_{AB},\xi_{C})=\int_{1}^{\xi_C}\int_{-1}^{+1}(\xi+\nu)^{n}(\xi-\nu)^{n'}\\
\times e^{-\xi\left[\frac{1}{2}\left(\zeta+\zeta' \right)R_{AB} \right]-\nu\left[\frac{1}{2}\left(\zeta-\zeta' \right)R_{AB} \right]}\\
\times\overline{\mathcal{P}}_{lm } \left( \frac{1+\xi\nu}{\xi+\nu}\right)\overline{\mathcal{P}}_{l'm'} \left( \frac{1-\xi\nu}{\xi-\nu}\right)\\
\times P_{L}^{\vert M \vert}\left(\xi \right)\overline{\mathcal{P}}_{LM}\left(\nu \right)d\xi d\nu
\end{multline}
\begin{multline}\label{eq:KGENAUXILIARY}
\mathcal{K}_{nlm,n'l'm'}^{LM}(\zeta,\zeta',R_{AB},\xi_{C})=\int_{\xi_C}^{\infty}\int_{-1}^{+1}(\xi+\nu)^{n}(\xi-\nu)^{n'}\\
\times e^{-\xi\left[\frac{1}{2}\left(\zeta+\zeta' \right)R_{AB} \right]-\nu\left[\frac{1}{2}\left(\zeta-\zeta' \right)R_{AB} \right]}\\
\times\overline{\mathcal{P}}_{lm } \left( \frac{1+\xi\nu}{\xi+\nu}\right)\overline{\mathcal{P}}_{l'm'} \left( \frac{1-\xi\nu}{\xi-\nu}\right)\\
\times Q_{L}^{\vert M \vert}\left(\xi \right)\overline{\mathcal{P}}_{LM}\left(\nu \right)d\xi d\nu,
\end{multline}
where, $\overline{\mathcal{P}}_{l \vert m \vert}(x)$ are the normalized associated Legendre functions. 

The auxiliary functions $\mathcal{J}^{LM}, \mathcal{K}^{LM}$ can be defined in a simpler form through relation given for product of two normalized associated Legendre functions centered on points $A$, $B$ in prolate spheroidal coordinates \cite{Guseinov1976},
\begin{multline}\label{eq:PTWOANLEGENDRE}
\left[(\xi^2-1)(1-\nu^2) \right]^{\Lambda-\frac{\lambda+\lambda'}{2}}\overline{\mathcal{P}}_{l\lambda }(\cos \theta _{A}) \overline{\mathcal{P}}_{l'\lambda'}(\cos \theta _{B})\\
=\sum _{\alpha =-(2\Lambda-\lambda)}^{l}\sum_{\beta =\lambda'}^{l'}
\sum _{q=0}^{\alpha +\beta-2\Lambda-\lambda-\lambda'}{g_{\alpha \beta}^{q}(l\lambda ,l'\lambda';\Lambda)}\\
\times{\left[\frac{({\xi \nu })^{q}}
{(\xi +\nu)^{\alpha }(\xi -\nu )^{\beta }}\right]}.
\end{multline}
which is obtained in explicit form by taking advantage of binomial expansion theorem,
\begin{multline} \label{eq:BINOMXP}
(x+a)^{N_{1}}(x-a)^{N_2} \\
=\sum_{s=0}^{N_1+N_2} F_{s}(N_{1},N_{2})x^{N_{1}+N_{2}-s}a^{s},
\end{multline} 
with,
\begin{align*}
\cos \theta_{A}=\frac{1+\xi \nu }{\xi +\nu };\hspace{5mm}\cos \theta _{B}=\frac{1-\xi \nu }{\xi -\nu }.
\end{align*}
The coefficients $g_{\alpha\beta}^{q}$ occurring in Eq. (\ref{eq:PTWOANLEGENDRE}) are determined by,
\begin{align} \label{eq:GABC1}
g_{\alpha\beta}^{q}(l\lambda,l'\lambda)=g_{\alpha\beta}^{0}(l\lambda,l'\lambda)F_{q}(\alpha+\lambda,\beta-\lambda)
\end{align}
\begin{align} \label{eq:GABC1}
g_{\alpha\beta}^{0}(l\lambda,l'\lambda)=\sum_{s=0}^{\lambda}(-1)^{s}F_{s}(\lambda)D_{\alpha+2\lambda-2s}^{l\lambda}D_{\beta}^{l'\lambda},
\end{align}
\begin{multline} \label{eq:GAUNT}
D_{\beta}^{l\lambda}=\frac{1}{2^l}(-1)^{(l-\beta)/2}\left[\frac{2l+1}{2}\frac{F_{l}(l+\lambda)}{F_{\lambda}(l)} \right]^{1/2}\\
\times F_{(l-\beta)/2}(l)F_{\beta-\lambda}(l+\beta),
\end{multline}
where, $\lambda= \vert m \vert$, $\lambda'= \vert m' \vert$ and the quantities $F_{s}(N,N')$ are the generalized binomial coefficients. They are given as,
\begin{align} \label{eq:GENBINOM}
F_{s}(N,N')=\sum_{s'}(-1)^{s'}F_{s-s'}(N)F_{s'}(N')
\end{align}
with, $\frac{1}{2}\left[(s-N)+\vert s-N \vert \right]\leq s' \leq min(s,N)$ and $F_{s}(N)$ are binomial coefficients indexed by $N$ and $s$ is usually written $\left(\begin{array}{cc}N\\s\end{array} \right)$, respectively.

The Eqs. (\ref{eq:JGENAUXILIARY}, \ref{eq:KGENAUXILIARY}) are, therefore, obtained as follows,
\begin{multline}\label{eq:JKGENAUXILIARY}
\left[\begin{array}{cc}
   \mathcal{J}^{LM}_{nlm,n'l'm'}\left(\zeta,\zeta',R_{AB},\xi_{C} \right)
   \\
  \mathcal{K}^{LM}_{nlm,n'l'm'}\left(\zeta,\zeta',R_{AB},\xi_{C} \right)
\end{array} \right]=\sum_{\alpha\beta q}g_{\alpha \beta}^{q}(l\lambda ,l'\lambda';\Lambda)\\
\times \left[\begin{array}{cc}
   \mathcal{J}^{LM,q}_{n-\alpha, n'-\beta}\left(\zeta,\zeta',R_{AB},\xi_{C} \right)
   \\
  \mathcal{K}^{LM,q}_{n-\alpha, n'-\beta}\left(\zeta,\zeta',R_{AB},\xi_{C} \right)
\end{array} \right],
\end{multline}
with,
\begin{multline}\label{eq:JKREDAUXILIARY}
\left[\begin{array}{cc}
   \mathcal{J}^{L\Lambda,q}_{n-\alpha,n'-\beta}\left(\zeta,\zeta',R_{AB},\xi_{C} \right)
   \\
  \mathcal{K}^{L\Lambda,q}_{n-\alpha,n'-\beta}\left(\zeta,\zeta',R_{AB},\xi_{C} \right)
\end{array} \right]\\
=\int_{{\tiny \left[\begin{array}{cc} 1 \\ \xi_{C} \end{array} \right]}}^{{\tiny \left[\begin{array}{cc} \xi_{C} \\ \infty \end{array} \right]}}\int_{-1}^{1}{\left(\xi\nu \right)^{q}\left(\xi+\nu \right)^{n-\alpha}\left(\xi-\nu \right)^{n'-\beta}}\\
\times e^{-\xi\left[\frac{1}{2}\left(\zeta+\zeta' \right)R_{AB} \right]-\nu\left[\frac{1}{2}\left(\zeta-\zeta' \right)R_{AB} \right]}\\
\times \begin{bmatrix}
\mathcal{P}_{L}^{\vert \Lambda \vert} \left(\xi\right) \\ \mathcal{Q}_{L}^{\vert \Lambda \vert} \left(\xi\right)]
\end{bmatrix}\overline{\mathcal{P}}_{L \vert \Lambda \vert}\left(\nu \right)d\xi d\nu.
\end{multline}
There are no known convergent series representation formulas, free from specifications on parameters for power functions such as $(\xi+\nu)^{N_{1}}, (\xi-\nu)^{N_{2}}$, $\left\lbrace N_{1}, N_{2} \right\rbrace \in \mathbb{R}$, yet and that poses an obstacle to analytically reduce the $\mathcal{J}^{L\Lambda,q}$, $\mathcal{K}^{L\Lambda,q}$ auxiliary functions to one variable $w_{\mu}^{q}, L_{\mu}^{q}, k_{\mu}^{q}$ auxiliary functions introduced in \cite{Harris2002}. Thus, the solution should be obtained on the basis of numerical methods. Note that, taking advantage of binomial expansion method for terms containing the angular part of Slater-type orbitals in order to simplify the expressions increases the number of integrals should be numerically calculated. In {\sl Mathematica} programming language instead of using Eq. (\ref{eq:JKGENAUXILIARY}), direct computation of Eqs. (\ref{eq:JGENAUXILIARY}, \ref{eq:KGENAUXILIARY}) are faster. The given relations for auxiliary functions in Eq. (\ref{eq:JKREDAUXILIARY}) are calculated by using different expressions of Legendre polynomials and compared according to computational time in Fig. (\ref{fig:CPUT}) as sample. The discussions on results are made in the next section.

\section{\label{sec:ResDis}Results and discussions}

\begin{figure}[!t]
\centering
\includegraphics[width=0.50\textwidth,height=0.22\textheight]{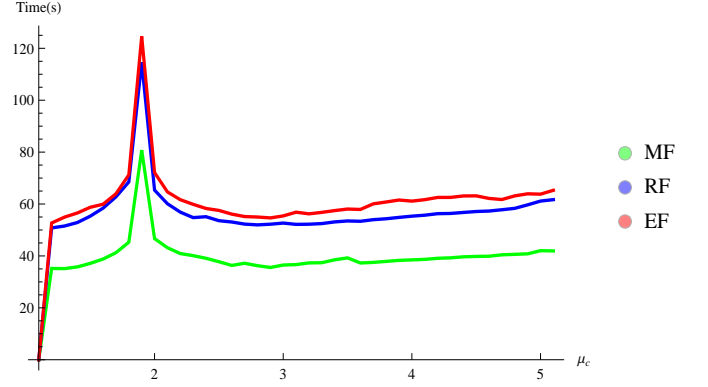}
\caption{\label{fig:CPUT} CPU time for computation of $\mathcal{J}^{L\Lambda,q}_{N_1N_2}$ auxiliary function in Eq.(\ref{eq:JKREDAUXILIARY}) according to methods used for calculation of the Legendre polynomials, where, {\sl Mathematica} built-function (MF), recurrence relation formula (RF), explicit formula (EF) and, $L=3$, $\Lambda=1$, $q=0$, $N_{1}=3$, $N_{2}=2$, $p_{1}=2.5$, $p_{2}=1.5$.}
\end{figure}

The literature currently, lack of benchmark values for multi$-$center integrals when Slater-type orbitals are used. Recently, a robust numerical Global$-$adaptive strategy with Gauss$-$Kronrod extension has been applied for two$-$center integrals through prolate spheroidal coordinates and fourier transform method in \cite{Bagci2014, Bagci2015, Silverstone2014}. Benchmark results have been presented for them. The hermitian properties are thus, represented correctly free from specification on quantum numbers, orbital parameters and internuclear distances. In this study it is extended for solution of three$-$center integrals. The algorithm described in \cite{Bagci2014} has been incorporated into a computer program written in the {\sl Mathematica} programming language with the included numerical computation packages for solving Eqs. (\ref{eq:METHREECENTER}, \ref{eq:JGENAUXILIARY}, \ref{eq:KGENAUXILIARY}). The {\sl Mathematica} programming language can handle approximate real numbers with any number of digits and it is suitable for benchmark evaluation. It is also provides a uniquely integrated and automated environment for parallel computing. It is allows us to compute the formulas including summations using all cores of PC effectively via \textbf{ParallelSum} command instead of \textbf{Sum}. Note that, in this study all results are given in atomic units (a.u.).
\begin{table}[!htp]
\caption{\label{tab:Spectrum1} The values for auxiliary functions defined in Eq.(\ref{eq:JKREDAUXILIARY}) with $p_{1}=\frac{1}{2}\left(\zeta+\zeta' \right)R_{AB}$, $p_{2}=\frac{1}{2}\left(\zeta-\zeta' \right)R_{AB}$ and $N_{1},N_{2} \in \mathbb{R}^{+}$.}
\begin{ruledtabular}
\begin{tabular}{cccccccc}
$L$ & $\Lambda$ & $q$ & $N_{1}$ & $N_{1}$ & $p_{1}$ & $p_{2}$ & Results
\\
\hline
$0$ & $0$ & $0$ & $1.0$ & $1.0$ & $1.5$ & $1.5$ & 
\begin{tabular}[c]{@{}l@{}}3.62319 79582 17897 45490 E$-$01\\1.75859 65139 47296 72718 E$-$01\end{tabular}
\\
\hline
$1$ & $0$ & $2$ & $3.0$ & $2.0$ & $4.5$ & $0.5$ & 
\begin{tabular}[c]{@{}l@{}}1.02307 57195 13525 65648 E$-$03\\1.20185 35031 71549 79713 E$-$03 \end{tabular}
\\
\hline
$2$ & $2$ & $2$ & $3.0$ & $2.0$ & $4.5$ & $0.5$ & 
\begin{tabular}[c]{@{}l@{}}1.26482 46553 60927 73809 E$-$02\\3.05572 36905 83528 19812 E$-$04\end{tabular}
\\
\hline
$3$ & $2$ & $5$ & $9.0$ & $4.0$ & $22.5$ & $0.1$ & 
\begin{tabular}[c]{@{}l@{}}1.54156 95966 91532 03328 E$-$11\\9.77333 04203 13413 96225 E$-$18\end{tabular}
\\
\hline
$5$ & $4$ & $6$ & $15.0$ & $3.0$ & $4.0$ & $0.1$ & 
\begin{tabular}[c]{@{}l@{}}7.45864 92397 67729 51682 E$+$06\\1.65225 27526 05586 45258 E$+$05\end{tabular}
\\
\hline
$0$ & $0$ & $0$ & $1.2$ & $1.5$ & $1.5$ & $1.5$ & 
\begin{tabular}[c]{@{}l@{}}4.93516 66112 08595 80377 E$-$01\\3.72444 04752 38238 19870 E$-$01\end{tabular}
\\
\hline
$1$ & $0$ & $2$ & $3.3$ & $2.4$ & $4.5$ & $0.5$ & 
\begin{tabular}[c]{@{}l@{}}7.02522 66592 59862 42272 E$-$04\\4.07093 04267 92761 66995 E$-$05\end{tabular}
\\
\hline
$2$ & $2$ & $2$ & $3.5$ & $2.5$ & $4.5$ & $0.5$ & 
\begin{tabular}[c]{@{}l@{}}2.01392 20090 23026 29191 E$-$02\\6.86518 34228 55977 92648 E$-$04\end{tabular}
\\
\hline
$3$ & $2$ & $5$ & $9.9$ & $4.1$ & $22.5$ & $0.1$ & 
\begin{tabular}[c]{@{}l@{}}2.74437 86677 61624 01320 E$-$11\\2.53854 67628 59329 46820 E$-$17\end{tabular}
\\
\hline
$5$ & $4$ & $6$ & $15.2$ & $3.6$ & $4.0$ & $0.1$ & 
\begin{tabular}[c]{@{}l@{}}9.75947 26622 39909 40431 E$+$06\\4.76137 54661 40175 09867 E$+$05\end{tabular}
\\
\hline
$5$ & $4$ & $6$ & $15.2$ & $3.6$ & $0.1$ & $4.0$ & 
\begin{tabular}[c]{@{}l@{}}9.01557 46105 79752 64049 E$+$08\\9.49361 57667 79094 34667 E$+$37\end{tabular}
\\
\hline
$5$ & $4$ & $6$ & $15.2$ & $3.6$ & $0.1$ & $4.0$ & 
\begin{tabular}[c]{@{}l@{}}6.67509 82662 40505 33704 E$+$08\\1.34769 33998 41265 16644 E$+$36\end{tabular}
\end{tabular}
\end{ruledtabular}
\end{table}
The calculation results are presented in Tables \ref{tab:Spectrum1}$-$\ref{tab:Threecenter4} and Fig.\ref{fig:CPUT} for arbitrary values of quantum numbers, orbital parameters and internuclear distances. The comparisons are made with expansion methods which are given for expansion of wave function in Eq. (\ref{eq:WAVEEXPTHEO}) and for charge density expansion to same center by following formula \cite{Guseinov2000, Guseinov2002-1, Guseinov2007}
\begin{multline} \label{eq:CHARSEXPTHEO}
\rho_{nlm,n'l'm'}(\zeta,\vec r;\zeta',\vec r)\\
=\sum_{l''=\vert l-l'\vert}^{l+l'}\sum_{m''=-l''}^{l''} W_{nlm,n'l'm',n+n'-1l''m''}(\zeta,\zeta',z)\\
\times\chi_{n+n'-1l''m''}(z,\vec r).
\end{multline}
They are useful to reduce the three$-$center integrals to basic nuclear attraction integrals, 
\begin{align} \label{eq:BASICNUCATTRACT}
J_{\kappa\lambda\tau}(z,\vec{R}_{BC})=\frac{1}{\sqrt{4\pi}}\int \chi_{\kappa\lambda\tau}^{*}(z,\vec{r}_{B})\frac{1}{r_{C}}dv_{1}.
\end{align}
The Eqs. (\ref{eq:WAVEEXPTHEO}, \ref{eq:CHARSEXPTHEO}) are used to transform the wave function centered at $A$ to a wave function centered at $B$ then, to transform the charge density centered on same positions to a single wave function, respectively. Here, $z=\zeta+\zeta'$, $\vec{R}_{BC}=\vec{AC}$. The resulting basic nuclear attraction integrals are calculated by the following formula, \cite{Guseinov2002-2},
\begin{multline} \label{eq:BASICNUCATTRACTEVAL}
J_{\kappa\lambda\tau}(z,\vec{R})=\\
\frac{2^{\kappa}}{2\lambda+1}\sqrt{\frac{2}{z}}\frac{\Gamma(\kappa+\lambda+2)}{\sqrt{\Gamma(2\kappa+1)}}\frac{1}{(zR)^{\lambda+1}}\\
\times \left(1-\frac{\Gamma(\kappa+\lambda+2,zR)}{\Gamma(\kappa+\lambda+2)}+\frac{(zR)^{2\lambda+1}\Gamma(\kappa-\lambda+1,zR)}{\Gamma(\kappa+\lambda+2)} \right)\\
\times Y_{\lambda,\tau}(\theta,\phi),
\end{multline}
where, $\Gamma(n,m)$ are incomplete gamma functions \cite{Abramowitz1972}. 

In Fig. \ref{fig:CPUT} the Eq. (\ref{eq:JKREDAUXILIARY}) is investigated according to computational time in {\sl Mathematica} programming language. {\sl Mathematica} includes all the common special functions of mathematical physics. It also provides easy way of computing them precisely. Here, explicit formula (EF) \cite{Guseinov1995},
\begin{align}\label{eq:ANALLEG}
\overline{\mathcal{P}}_{l\lambda}(x)=\left(1-x^2 \right)^{\frac{\lambda}{2}}\sum_{k}b_{l\lambda}^{k}x^{l-\lambda-2k},
\end{align}
\begin{multline}\label{eq:ANALLEGCOFF}
b_{l\lambda}^{k}=\frac{1}{2^l}\left[\frac{2l+1}{2F_{\lambda}(l)F_{\lambda}(l+\lambda)} \right]^{\frac{1}{2}}\\
\times(-1)^{k}F_{k}(\lambda+k)F_{l-k}(2l-2k)F_{l-\lambda-2k}(l-k),
\end{multline}
where, $0 \leq k \leq E\left[\frac{l-\lambda}{2} \right]$, recurrence relation formula (RF) \cite{NIST2014} of Legendre polynomials are compared with {\sl Mathematica} built-function (MF) \textbf{LegendreP[n,m,x]}. They are presented with red, blue, green lines in Fig. \ref{fig:CPUT}, respectively. It can be seen from this figure, the direct use of {\sl Mathematica} buit-function, given for computing of Legendre polynomials, in numerical integration of Eq. (\ref{eq:JKREDAUXILIARY}) provide the results faster then explicit or recurrence relation formulas. \\
The results for calculation of Eq. (\ref{eq:JKREDAUXILIARY}) are also presented in Table \ref{tab:Spectrum1} with integer and noninteger values of principal quantum numbers. The first, second rows are obtained from calculation $\mathcal{J}^{L\Lambda,q}_{N_1N_2}$ and $\mathcal{K}^{L\Lambda,q}_{N_1N_2}$ functions, respectively. The auxiliary functions $w_{\mu}^{q}, L_{\mu}^{q}, k_{\mu}^{q}$ defined in \cite{Harris2002} for three-center integrals are special case of Eq. (\ref{eq:JKREDAUXILIARY}). Hence, we believe an importance of present the results for general form of Eq. (\ref{eq:JKREDAUXILIARY}).
\\
The results for Eqs. (\ref{eq:JGENAUXILIARY}, \ref{eq:KGENAUXILIARY}) and Eq. (\ref{eq:JKGENAUXILIARY}) are presented in Tables \ref{tab:Spectrum2}, \ref{tab:Spectrum3}. They are given in first, second rows for $\mathcal{J}^{LM}_{nlm,n'l'm'}$ and $ \mathcal{K}^{LM}_{nlm,n'l'm'}$ auxiliary functions, respectively. Note that, the Eq. (\ref{eq:JKGENAUXILIARY}) is only differ from Eqs. (\ref{eq:JGENAUXILIARY}, \ref{eq:KGENAUXILIARY}) in that, the normalized associated Legendre polynomials on right-hand side are expanded via Eq. (\ref{eq:PTWOANLEGENDRE}) and the numerical global adaptive method is performed to remaining parts. Performing the calculations in {\sl Mathematica} programming language for such formulas contain summations is disadvantageous in terms of calculation time. The results in Tables \ref{tab:Spectrum2}, \ref{tab:Spectrum3} shows that, numerical Global adaptive method with Gauss-Kronrod extension can be used for computation of Eqs. (\ref{eq:JGENAUXILIARY}, \ref{eq:KGENAUXILIARY}) which is eliminate necessity applying binomial expansion theorem. 

In Tables \ref{tab:Threecenter1}, \ref{tab:Threecenter2}, \ref{tab:Threecenter3} the results obtained for three-center integrals are presented for upper limit of summation $L$ is $L=30$. The correct digits are underlined. The digits in bold indicate the convergence property of used method. In first and second rows benchmark results obtained from numerical global adaptive method and results those found in the literature are presented, respectively. Later rows are the results obtained from expansion of the wave-function method and they are in complete agreement with ones from E. \c{S}ahin (personal communication), where the upper imit of summations $N_{e}$ are given in parenthesis. The expansion of the wave-function method is tested up to upper limit of summation $N_{e}$, $N_{e}=160$. It is observed that, the results hardly convergent with 10$-$digits for given quantum numbers and orbital parameters in table \ref{tab:Threecenter2}. In other tables the convergence remains between 5$-$digits and 10$-$digits. Note that, it is necessary to take into account eight summation and four of them should be infinite in expansion of the wave-function method if NSTOs are used. The values presented in Tables \ref{tab:Threecenter1}, \ref{tab:Threecenter2}, \ref{tab:Threecenter3} for STOs clearly demonstrate pointlessness of such an attempt.\\
On the other hand in our previous papers \cite{Bagci2014,Bagci2015} it have been proved that the numerical Global adaptive method with Gauss-Kronrod extension is able to give benchmark values. In particular for Table \ref{tab:Threecenter4} the results are presented for different upper limit of summation $L$ appears in Eq. (\ref{eq:METHREECENTER}). They differs from upper limit of summation $N_{e}$ used in expansion of STOs in that they are given in brackets. Convergence property of Eq. (\ref{eq:METHREECENTER}) is examined in this table. It is found that, by increasing the upper limit of summation the results are convergence to exact values. The results obtained up to upper limit of summation $L$ is $L=40$. It is achieved to 25$-$digits accuracy by determining the upper limit of summation $L$ is $L=30$ accordingly, there is no necessity of performing calculations with upper limit of summation higher than $L=30$ unless more precise results required for a given values of parameters. It should be point out that, the summation appears in Eq. (\ref{eq:METHREECENTER}) should not be regarded as having same characteristic with summation arising in expansion of STOs. It is based on expansion of spherical harmonics which have form a complete set of orthonormal functions. Any square-integrable function can be expanded as a linear combination of spherical harmonics. The convergence problems arising in expansion NSTOs can not exist in our method. Without any computational difficulty by increasing the upper limit of summation $L$ can be achieved to desired accuracy rapidly.

\section*{Acknowledgement}
A.B. acknowledges funding for a postdoctoral research fellowship from innov@pole: the Auvergne Region and FEDER.

\clearpage
\begin{widetext}
\begin{table*}[!htb]
\caption{\label{tab:Spectrum2} The values for auxiliary functions defined in Eqs. (\ref{eq:JGENAUXILIARY},\ref{eq:KGENAUXILIARY}) and Eq. (\ref{eq:JKGENAUXILIARY}) with $p_{1}=\frac{1}{2}\left(\zeta+\zeta' \right)R_{AB}$, $p_{2}=\frac{1}{2}\left(\zeta-\zeta' \right)R_{AB}$ and $n,n' \in \mathbb{N}^{+}$.}
\begin{ruledtabular}
\begin{tabular}{ccccccccccc}
$L$ & $M$ & $n$ & $l$ & $m$ & $n'$ & $l'$ & $m'$ & $p_{1}$ & $p_{2}$ & Eqs. (\ref{eq:JGENAUXILIARY},\ref{eq:KGENAUXILIARY}) and Eq. (\ref{eq:JKGENAUXILIARY})
\\
\hline
$0$ & $0$ & $1.0$ & $0$ & $0$ & $1.0$ & $0$ & $0$ & $1.5$ & $2.5$ & 
\begin{tabular}[c]{@{}l@{}}2.90399 07988 62162 00938 68706 03684 E$-$01\\1.47689 95097 54804 95414 54214 62847 
E$-$01\end{tabular}
\\
\hline
$2$ & $0$ & $3.0$ & $0$ & $0$ & $2.0$ & $0$ & $0$ & $0.01$ & $2.5$ & 
\begin{tabular}[c]{@{}l@{}}3.42865 83511 73345 68718 16357 54166 E$-$00\\8.97033 20343 34376 52929 54799 07851 E$+$05\end{tabular}
\\
\hline
$5$ & $0$ & $9.0$ & $0$ & $0$ & $4.0$ & $0$ & $0$ & $22.5$ & $0.1$ & 
\begin{tabular}[c]{@{}l@{}}6.61192 17510 24844 44964 35310 16794 E$-$12\\1.27733 49592 55934 35439 28200 66327 E$-$23\end{tabular}
\\
\hline
$10$ & $0$ & $12.0$ & $0$ & $0$ & $8.0$ & $0$ & $0$ & $40.0$ & $0.001$ & 
\begin{tabular}[c]{@{}l@{}}1.73471 85107 61691 61997 46481 63897 E$-$20\\1.87470 00384 84104 39018 32142 95122 E$-$41\end{tabular}
\\
\hline
$1$ & $0$ & $3.0$ & $1$ & $0$ & $2.0$ & $1$ & $0$ & $2.3$ & $4.5$ & 
\begin{tabular}[c]{@{}l@{}}1.06208 64400 83910 27727 15664 09620 E$-$00\\2.45315 16908 16648 21895 84639 84955 E$-$01\end{tabular}
\\
\hline
$6$ & $0$ & $6.0$ & $4$ & $0$ & $5.0$ & $3$ & $0$ & $0.001$ & $0.001$ & 
\begin{tabular}[c]{@{}l@{}}2.25821 42942 16571 75095 60318 57985 E$+$46\\2.48928 81670 28192 48883 41015 71291 E$+$56\end{tabular}
\\
\hline
$1$ & $0$ & $2.0$ & $1$ & $1$ & $2.0$ & $1$ & $1$ & $0.01$ & $0.01$ & 
\begin{tabular}[c]{@{}l@{}}1.39224 44783 13768 61934 60782 72901 E$-$02\\1.63293 63668 41775 85565 65259 43851 E$+$03\end{tabular}
\\
\hline
$3$ & $0$ & $6.0$ & $4$ & $3$ & $5.0$ & $3$ & $3$ & $8.5$ & $0.1$ & 
\begin{tabular}[c]{@{}l@{}}1.22009 58147 71182 05449 05454 44137 E$-$04\\1.63709 36934 93917 83173 07173 88885 E$-$08\end{tabular}
\\
\hline
$3$ & $2$ & $5.0$ & $3$ & $3$ & $3.0$ & $2$ & $1$ & $1.5$ & $1.5$ & 
\begin{tabular}[c]{@{}l@{}}8.36879 35046 74242 89056 27574 95673 E$+$01\\2.18611 99801 01738 31430 63739 74869 E$-$00\end{tabular}
\\
\hline
$3$ & $2$ & $10.0$ & $4$ & $4$ & $8.0$ & $2$ & $2$ & $0.1$ & $9.0$ & 
\begin{tabular}[c]{@{}l@{}}8.73612 92750 76127 55419 06451 97990 E$+$06\\1.89278 99267 51317 93832 38020 79768 E$+$06\end{tabular}
\end{tabular}
\end{ruledtabular}
\end{table*} 
\begin{table*}
\caption{\label{tab:Spectrum3} The values for auxiliary functions defined in Eqs. (\ref{eq:JGENAUXILIARY},\ref{eq:KGENAUXILIARY}) and Eq. (\ref{eq:JKGENAUXILIARY}) with $p_{1}=\frac{1}{2}\left(\zeta+\zeta' \right)R_{AB}$, $p_{2}=\frac{1}{2}\left(\zeta-\zeta' \right)R_{AB}$ and $n,n' \in \mathbb{R}^{+}$.}
\begin{ruledtabular}
\begin{tabular}{ccccccccccc}
$L$ & $M$ & $n$ & $l$ & $m$ & $n'$ & $l'$ & $m'$ & $p_{1}$ & $p_{2}$ & Eq.(16,17) and Eq.(19)
\\
\hline
$0$ & $0$ & $1.1$ & $0$ & $0$ & $1.5$ & $0$ & $0$ & $1.5$ & $1.5$ & 
\begin{tabular}[c]{@{}l@{}}4.09565 63231 31448 87039 55147 40491 E$-$01\\2.93614 23789 04133 04860 14608 03840 E$-$01\end{tabular}
\\
\hline
$1$ & $0$ & $1.3$ & $0$ & $0$ & $1.8$ & $0$ & $0$ & $0.5$ & $0.1$ & 
\begin{tabular}[c]{@{}l@{}}3.33243 32418 64279 67130 02542 67866 E$-$01\\1.02388 22819 29861 91096 69560 06751 E$-$02\end{tabular}
\\
\hline
$2$ & $0$ & $3.2$ & $0$ & $0$ & $2.4$ & $0$ & $0$ & $0.01$ & $2.5$ & 
\begin{tabular}[c]{@{}l@{}}5.49415 76314 41209 10218 23695 48480 E$-$00\\7.89858 88444 51044 77570 43625 77546 E$+$06\end{tabular}
\\
\hline
$5$ & $0$ & $9.5$ & $0$ & $0$ & $4.8$ & $0$ & $0$ & $22.5$ & $0.1$ & 
\begin{tabular}[c]{@{}l@{}}8.85039 89771 63772 65903 44204 90792 E$-$12\\9.74443 14137 20230 92395 49657 75617 E$-$23\end{tabular}
\\
\hline
$10$ & $0$ & $12.5$ & $0$ & $0$ & $8.9$ & $0$ & $0$ & $40.0$ & $0.001$ & 
\begin{tabular}[c]{@{}l@{}}1.17276 22934 30019 21937 27274 20564 E$-$20\\5.51171 88865 58272 19013 29428 46796 E$-$41\end{tabular}
\\
\hline
$1$ & $0$ & $3.3$ & $1$ & $0$ & $2.6$ & $1$ & $0$ & $2.3$ & $4.5$ & 
\begin{tabular}[c]{@{}l@{}}1.87161 11340 32480 14502 70364 65188 E$-$00\\6.65100 22118 49834 63694 23667 20358 E$-$01\end{tabular}
\\
\hline
$6$ & $0$ & $5.9$ & $4$ & $0$ & $6.1$ & $3$ & $0$ & $0.001$ & $0.001$ & 
\begin{tabular}[c]{@{}l@{}}7.38511 70805 47763 29724 30676 68341 E$+$46\\1.24500 66118 34410 00379 04215 62074 E$+$60\end{tabular}
\\
\hline
$1$ & $0$ & $2.3$ & $1$ & $1$ & $2.1$ & $1$ & $1$ & $0.01$ & $0.01$ & 
\begin{tabular}[c]{@{}l@{}}1.85417 21104 44330 08693 85023 65883 E$-$01\\1.40783 85008 42084 72060 75228 55754 E$+$04\end{tabular}
\\
\hline
$3$ & $0$ & $5.9$ & $4$ & $3$ & $5.1$ & $3$ & $3$ & $8.5$ & $0.1$ & 
\begin{tabular}[c]{@{}l@{}}1.13313 80933 64382 91558 05780 65710 E$-$04\\1.56774 69588 19924 39645 11990 75063 E$-$08\end{tabular}
\\
\hline
$3$ & $0$ & $8.8$ & $4$ & $3$ & $10.3$ & $3$ & $3$ & $0.01$ & $0.01$ & 
\begin{tabular}[c]{@{}l@{}}4.97973 96013 63878 34830 92006 86281 E$+$04\\7.95070 98832 40529 88293 62402 30270 E$+$42\end{tabular}
\\
\hline
$3$ & $2$ & $4.5$ & $3$ & $3$ & $3.5$ & $2$ & $1$ & $1.5$ & $1.5$ & 
\begin{tabular}[c]{@{}l@{}}1.12197 95987 12589 22596 64533 39225 E$+$02\\2.51548 44999 37750 08034 41787 62496 E$-$00\end{tabular}
\\
\hline
$3$ & $2$ & $10.5$ & $4$ & $4$ & $8.1$ & $2$ & $2$ & $0.1$ & $9.0$ & 
\begin{tabular}[c]{@{}l@{}}1.10758 39108 61744 40668 98531 33050 E$+$07\\1.32690 39575 82905 48879 15023 14344 E$+$29\end{tabular}
\end{tabular}
\end{ruledtabular}
\end{table*}
\begin{table*}
\caption{\label{tab:Threecenter1} The values for three-center integrals, where position of the nuleus A, B, C in cartesian coordinates $\left\lbrace X, Y,Z \right\rbrace$: A=$\left\lbrace 0,0,0 \right\rbrace$, B=$\left\lbrace 0,0,-2.0143 \right\rbrace$, C=$\left\lbrace 0,0,-4.1934 \right\rbrace$, respectively.}
\begin{ruledtabular}
\begin{tabular}{ccccccccc}
$n$ & $l$ & $m$ & $\zeta$ & $n'$ & $l'$ & $m'$ & $\zeta'$ & Results
\\
\hline
$1.0$ & $0$ & $0$ & $1.24$ & $1.0$ & $0$ & $0$ & $5.67$ & \begin{tabular}[c]{@{}l@{}}\underline{2.94549 60536 73751 14101 41604} E$-$02\\\underline{2.94549 6054} E$-$02 {\footnotemark[1]}\\ \underline{2.9}\textbf{23}67 19340 54421 10253 45246 E$-$02 (5) \\ \underline{2.9}\textbf{54}20 14999 28792 85854 63190 E$-$02 (10)\\ \underline{2.945}\textbf{69} 30497 61154 50027 23145 E$-$02 (20)\\ \underline{2.94549} \textbf{53}848 92871 61006 54907 E$-$02 (40)\\ \underline{2.94549} \textbf{58}796 89501 79906 76961 E$-$02 (80)\\ \underline{2.94549 6053}\textbf{2} 13114 00159 34423 E$-$02 (160)\end{tabular}
\\
\hline
$1.1$ & $0$ & $0$ & $1.24$ & $1.1$ & $0$ & $0$ & $5.67$ & \begin{tabular}[c]{@{}l@{}}\underline{3.34089 64668 60903 49505 70936} E$-$02\hspace{0.85cm} \end{tabular}
\\
\hline
$1.2$ & $0$ & $0$ & $1.24$ & $1.2$ & $0$ & $0$ & $5.67$ & \begin{tabular}[c]{@{}l@{}}\underline{3.74811 24252 41741 71875 67668} E$-$02\hspace{0.85cm} \end{tabular}
\\
\hline
$1.3$ & $0$ & $0$ & $1.24$ & $1.3$ & $0$ & $0$ & $5.67$ & \begin{tabular}[c]{@{}l@{}}\underline{4.16291 70512 36793 53772 11297} E$-$02\hspace{0.85cm} \end{tabular}
\\
\hline
$1.0$ & $0$ & $0$ & $1.24$ & $2.0$ & $0$ & $0$ & $1.61$ & \begin{tabular}[c]{@{}l@{}}\underline{1.60664 60408 09377 56494 10579} E$-$01\\\underline{1.60664 60}78 E$-$01{\footnotemark[1]}\\\underline{1}.\textbf{57}770 82722 80008 33479 04172 E$-$01 (5)\\\underline{1.60}\textbf{08}4 83584 29304 06752 43872 E$-$01 (10)\\\underline{1.60}\textbf{75}3 33306 97032 54324 80156 E$-$01 (20)\\\underline{1.606}\textbf{73} 68838 68511 26148 86494 E$-$01 (40)\\\underline{1.606}\textbf{59} 82475 67392 07711 26316 E$-$01 (80)\\\underline{1.60664} \textbf{10}673 25488 09376 70768 E$-$01 (160)\end{tabular}
\\
\hline
$1.1$ & $0$ & $0$ & $1.24$ & $2.1$ & $0$ & $0$ & $1.61$ & \begin{tabular}[c]{@{}l@{}}\underline{1.69763 82691 19573 36045 53004} E$-$01\hspace{0.85cm} \end{tabular}
\\
\hline
$1.2$ & $0$ & $0$ & $1.24$ & $2.2$ & $0$ & $0$ & $1.61$ & \begin{tabular}[c]{@{}l@{}}\underline{1.78213 08867 60019 08900 15372} E$-$01\hspace{0.85cm} \end{tabular}
\\
\hline
$1.3$ & $0$ & $0$ & $1.24$ & $2.3$ & $0$ & $0$ & $1.61$ & \begin{tabular}[c]{@{}l@{}}\underline{1.86000 31133 33942 68781 62339} E$-$01\hspace{0.75cm} \end{tabular}
\footnotetext[1]{\cite{Bouferguene1998}}
\end{tabular}
\end{ruledtabular}
\end{table*}
\begin{table*}

\caption{\label{tab:Threecenter2} The values for three-center integrals, where $\zeta=\zeta'=2.0$; position of the nuleus A, B, C in cartesian coordinates $\left\lbrace X, Y,Z \right\rbrace$: A=$\left\lbrace 0,0,0 \right\rbrace$, B=$\left\lbrace 0,0,6 \right\rbrace$, C=$\left\lbrace 0,0,-7 \right\rbrace$, respectively.}
\begin{ruledtabular}
\begin{tabular}{ccccccc}
$n$ & $l$ & $m$ & $n'$ & $l'$ & $m'$ & Results
\\
\hline
$2.0$ & $0$ & $0$ & $2.0$ & $0$ & $0$ & \begin{tabular}[c]{@{}l@{}}\underline{4.53377 50011 42666 45050 53528} E$-$04\\\underline{4.53377 50011 426}5 E$-$04{\footnotemark[1]}\\\underline{4.53377 50011} 15138 130 E$-$04{\footnotemark[2]}\\\underline{4}.\textbf{49}978 10366 11814 81692 67046 E$-$04 (5)\\\underline{4.5337}\textbf{1} 93448 02014 78421 78747 E$-$04 (10)\\\underline{4.53377 500}\textbf{06} 71569 59704 90361 E$-$04 (20)\\\underline{4.53377 5001}\textbf{3} 47449 39243 55321 E$-$04 (40)\\\underline{4.53377 50011} \textbf{98}380 66714 63012 E$-$04 (80)\\\underline{4.53377 50011} \textbf{56}735 48522 57686 E$-$04 (160)\end{tabular}
\\
\hline
$2.1$ & $0$ & $0$ & $2.1$ & $0$ & $0$ & \begin{tabular}[c]{@{}l@{}}\underline{5.51824 66271 98490 20314 29459} E$-$04\hspace{0.85cm} \end{tabular}
\\
\hline
$2.2$ & $0$ & $0$ & $2.2$ & $0$ & $0$ & \begin{tabular}[c]{@{}l@{}}\underline{6.66626 60479 98863 93114 16105} E$-$04\hspace{0.85cm} \end{tabular}
\\
\hline
$2.3$ & $0$ & $0$ & $2.3$ & $0$ & $0$ & \begin{tabular}[c]{@{}l@{}}\underline{7.99547 53772 28315 43332 78068} E$-$04\hspace{0.85cm} \end{tabular}
\\
\hline
$2.0$ & $1$ & $0$ & $2.0$ & $0$ & $0$ & \begin{tabular}[c]{@{}l@{}}\underline{6.49989 25189 10135 10589 99466} E$-$04 \\\underline{6.49989 25189 101}1 E$-$04{\footnotemark[1]}\hspace{2.7cm} \end{tabular}
\\
\hline
$2.1$ & $1$ & $0$ & $2.1$ & $0$ & $0$ & \begin{tabular}[c]{@{}l@{}}\underline{7.90609 20760 42277 16143 10272} E$-$04\hspace{0.85cm} \end{tabular}
\\
\hline
$2.2$ & $1$ & $0$ & $2.2$ & $0$ & $0$ & \begin{tabular}[c]{@{}l@{}}\underline{9.54293 49790 09494 09113 15449} E$-$04\hspace{0.85cm} \end{tabular}
\\
\hline
$2.3$ & $1$ & $0$ & $2.3$ & $0$ & $0$ & \begin{tabular}[c]{@{}l@{}}\underline{1.14342 80666 86023 44617 17750} E$-$03\hspace{0.85cm} \end{tabular}
\\
\hline
$2.0$ & $1$ & $0$ & $2.0$ & $1$ & $0$ & \begin{tabular}[c]{@{}l@{}}\underline{9.69666 68121 64802 10682 05157} E$-$04 \\\underline{9.69666 68121 64}77 E$-$04{\footnotemark[1]}\hspace{2.7cm} \end{tabular}
\\
\hline
$2.1$ & $1$ & $0$ & $2.1$ & $1$ & $0$ & \begin{tabular}[c]{@{}l@{}}\underline{1.17838 66960 60512 32340 53458} E$-$04\hspace{0.85cm} \end{tabular}
\\
\hline
$2.2$ & $1$ & $0$ & $2.2$ & $1$ & $0$ & \begin{tabular}[c]{@{}l@{}}\underline{1.42089 42554 34926 39305 88294} E$-$04\hspace{0.85cm} \end{tabular}
\\
\hline
$2.3$ & $1$ & $0$ & $2.3$ & $1$ & $0$ & \begin{tabular}[c]{@{}l@{}}\underline{1.70054 57496 59994 51520 55390} E$-$03\hspace{0.65cm} \end{tabular}
\footnotetext[1]{\cite{Rico1992}}
\footnotetext[2]{\cite{Peuker2008}}
\end{tabular}
\end{ruledtabular}
\end{table*}
\begin{table*}

\caption{\label{tab:Threecenter3} The values for three-center integrals, where $\zeta=\zeta'=1.6$; position of the nuleus A, B, C in cartesian coordinates $\left\lbrace X, Y,Z \right\rbrace$: A=$\left\lbrace 0,0,0 \right\rbrace$, B=$\left\lbrace 0,0,3 \right\rbrace$, C=$\left\lbrace 0,0,-3 \right\rbrace$, respectively.}
\begin{ruledtabular}
\begin{tabular}{ccccccc}
$n$ & $l$ & $m$ & $n'$ & $l'$ & $m'$ & Results
\\
\hline
$2.0$ & $0$ & $0$ & $2.0$ & $0$ & $0$ & \begin{tabular}[c]{@{}l@{}}\underline{7.41579 46662 21323 37855 75053} E$-$02\\\underline{7.41579 46662 2133} E$-$02{\footnotemark[1]}\\\underline{7.41579 46662 21323 37} E$-$02{\footnotemark[2]}\\\underline{7}.\textbf{39}466 24791 10067 86601 96990 E$-$02 (5)\\\underline{7.415}\textbf{95} 39598 23811 62963 42293 E$-$02 (10)\\\underline{7.4157}\textbf{4} 64572 83043 94945 93805 E$-$02 (20)\\\underline{7.4157}\textbf{8} 73240 73721 74843 59312 E$-$02 (40)\\\underline{7.41579} \textbf{22}349 71134 02044 81338 E$-$02 (80)\\\underline{7.41579} \textbf{51}272 77100 02097 74777 E$-$02 (160)\end{tabular}
\\
\hline
$2.1$ & $0$ & $0$ & $2.1$ & $0$ & $0$ & \begin{tabular}[c]{@{}l@{}}\underline{7.92564 04306 46454 51287 28675} E$-$02\hspace{0.85cm} \end{tabular}
\\
\hline
$2.2$ & $0$ & $0$ & $2.2$ & $0$ & $0$ & \begin{tabular}[c]{@{}l@{}}\underline{8.42445 84406 53803 96918 57249} E$-$02\hspace{0.85cm} \end{tabular}
\\
\hline
$2.3$ & $0$ & $0$ & $2.3$ & $0$ & $0$ & \begin{tabular}[c]{@{}l@{}}\underline{8.90946 29369 56705 18394 86907} E$-$02\hspace{0.85cm} \end{tabular}
\\
\hline
$2.0$ & $1$ & $0$ & $2.0$ & $0$ & $0$ & \begin{tabular}[c]{@{}l@{}}\underline{6.77544 93679 78440 06490 13878} E$-$02\\\underline{6.77544 93679 7845} E$-$02{\footnotemark[1]}\hspace{2.7cm} \end{tabular}
\\
\hline
$2.1$ & $1$ & $0$ & $2.1$ & $0$ & $0$ & \begin{tabular}[c]{@{}l@{}}\underline{7.10247 82121 12668 99509 25056} E$-$02\hspace{0.85cm} \end{tabular}
\\
\hline
$2.2$ & $1$ & $0$ & $2.2$ & $0$ & $0$ & \begin{tabular}[c]{@{}l@{}}\underline{7.39664 78499 31562 91249 65250} E$-$02\hspace{0.85cm} \end{tabular}
\\
\hline
$2.3$ & $1$ & $0$ & $2.3$ & $0$ & $0$ & \begin{tabular}[c]{@{}l@{}}\underline{7.65548 51830 78592 48565 84851} E$-$02\hspace{0.85cm} \end{tabular}
\\
\hline
$2.0$ & $1$ & $0$ & $2.0$ & $1$ & $0$ & \begin{tabular}[c]{@{}l@{}}\underline{6.75368 18945 49233 36334 86913} E$-$02\\\underline{6.75368 18945 492}6 E$-$02{\footnotemark[1]}\hspace{2.7cm} \end{tabular}
\\
\hline
$2.1$ & $1$ & $0$ & $2.1$ & $1$ & $0$ & \begin{tabular}[c]{@{}l@{}}\underline{6.91775 52123 72612 96334 96130} E$-$02\hspace{0.85cm} \end{tabular}
\\
\hline
$2.2$ & $1$ & $0$ & $2.2$ & $1$ & $0$ & \begin{tabular}[c]{@{}l@{}}\underline{7.02466 73478 35283 82859 84242} E$-$02\hspace{0.85cm} \end{tabular}
\\
\hline
$2.3$ & $1$ & $0$ & $2.3$ & $1$ & $0$ & \begin{tabular}[c]{@{}l@{}}\underline{7.07260 26896 46234 08137 55301} E$-$02\hspace{0.60cm} \end{tabular}
\footnotetext[1]{\cite{Rico1992}}
\footnotetext[2]{\cite{Peuker2008}}
\end{tabular}
\end{ruledtabular}
\end{table*}
\begin{table*}

\caption{\label{tab:Threecenter4} The values for three-center integrals, where $\zeta=3.6$, $\zeta'=1.6$; position of the nuleus A, B, C in cartesian coordinates $\left\lbrace X, Y,Z \right\rbrace$: A=$\left\lbrace 0,0,0 \right\rbrace$, B=$\left\lbrace 0,0,3 \right\rbrace$, C=$\left\lbrace 3,0,3 \right\rbrace$, respectively.}
\begin{ruledtabular}
\begin{tabular}{ccccccc}
$n$ & $l$ & $m$ & $n'$ & $l'$ & $m'$ & Results
\\
\hline
$2.0$ & $0$ & $0$ & $2.0$ & $0$ & $0$ & \begin{tabular}[c]{@{}l@{}}\underline{2}.\textbf{72}309 91701 60088 20662 71607 E$-$02 [1]\\\underline{2.7027}\textbf{6} 01923 98151 89286 78890 E$-$02 [5]\\\underline{2.70272 90}\textbf{21}9 79709 38269 45472 E$-$02 [10]\\\underline{2.70272 901}\textbf{87} 20151 76256 32122 E$-$02 [11]\\\underline{2.70272 901}\textbf{89} 88329 48522 85282 E$-$02 [12]\\\underline{2.70272 90190 09}\textbf{86}2 41008 43703 E$-$02 [15]\\\underline{2.70272 90190 09906 8540}\textbf{6} 45442 E$-$02 [20]\\\underline{2.70272 90190 09906 85401 14157} E$-$02 [30]\\\underline{2.70272 90190 09906 85401 14157} E$-$02 [40]\\\underline{2.70272 901}89 8 E$-$02{\footnotemark[1]}\\\underline{2.70272 90190} 2 E$-$02{\footnotemark[2]}\hspace{3.0cm} \end{tabular}
\\
\hline
$2.1$ & $0$ & $0$ & $2.1$ & $0$ & $0$ & \begin{tabular}[c]{@{}l@{}}\underline{2.99614 535}\textbf{94} 03007 59134 38353 E$-$02 [10]\\\underline{2.99614 53558 587}\textbf{18} 15620 66646 E$-$02 [15]\\\underline{2.99614 53558 58738 85669 00599} E$-$02 [30]\\\underline{2.99614 53558 58738 85669 00599} E$-$02 [40] \end{tabular}\hspace{0.05cm}
\\
\hline
$2.2$ & $0$ & $0$ & $2.2$ & $0$ & $0$ & \begin{tabular}[c]{@{}l@{}}\underline{3.30034 60}\textbf{72}4 25497 07263 17130 E$-$02 [10]\\\underline{3.30034 60682 949}\textbf{29} 39385 11310 E$-$02 [15]\\\underline{3.30034 60682 94915 87198 38}\textbf{70}0 E$-$02 [30]\\\underline{3.30034 60682 94915 87198 38698} E$-$02 [40] \end{tabular}\hspace{0.05cm}
\\
\hline
$2.3$ & $0$ & $0$ & $2.3$ & $0$ & $0$ & \begin{tabular}[c]{@{}l@{}}\underline{3.61354 341}\textbf{50} 29555 79331 66393 E$-$02 [10]\\\underline{3.61354 34103 210}\textbf{82} 87274 34304 E$-$02 [15]\\\underline{3.61354 34103 21023 85552 77843} E$-$02 [30]\\\underline{3.61354 34103 21023 85552 77839} E$-$02 [40] \end{tabular}\hspace{0.05cm}
\\
\hline
$2.0$ & $1$ & $0$ & $2.0$ & $0$ & $0$ & \begin{tabular}[c]{@{}l@{}}\underline{2.15396 7}\textbf{90}80 46322 59289 24910 E$-$02 [10]\\\underline{2.15396 78874 84}\textbf{62}6 90451 70130 E$-$02 [15]\\\underline{2.15396 78874 84896 61732 77914} E$-$02 [30]\\\underline{2.15396 78874 84896 61732 77914} E$-$02 [40]\\\underline{2.15396 7887}3 E$-$02{\footnotemark[1]} \end{tabular}\hspace{0.1cm}
\\
\hline
$2.1$ & $1$ & $0$ & $2.1$ & $0$ & $0$ & \begin{tabular}[c]{@{}l@{}}\underline{2.37228 0}\textbf{91}11 80051 67464 97075 E$-$02 [10]\\\underline{2.37228 08887 595}\textbf{65} 69293 67888 E$-$02 [15]\\\underline{2.37228 08887 59514 88542 814}\textbf{85} E$-$02 [30]\\\underline{2.37228 08887 59514 88542 81456} E$-$02 [40] \end{tabular}\hspace{0.1cm}
\\
\hline
$2.2$ & $1$ & $0$ & $2.2$ & $0$ & $0$ & \begin{tabular}[c]{@{}l@{}}\underline{2.59362 19}\textbf{46}2 42909 94539 77789 E$-$02 [10]\\\underline{2.59362 19225 73}\textbf{53}8 25384 87321 E$-$02 [15]\\\underline{2.59362 19225 73080 19834 6}\textbf{80}30 E$-$02 [30]\\\underline{2.59362 19225 73080 19834 67975} E$-$02 [40] \end{tabular}\hspace{0.1cm}
\\
\hline
$2.3$ & $1$ & $0$ & $2.3$ & $0$ & $0$ & \begin{tabular}[c]{@{}l@{}}\underline{2.81580 265}\textbf{27} 44329 75185 31334 E$-$02 [10]\\\underline{2.81580 26286 4}\textbf{70}36 51507 08634 E$-$02 [15]\\\underline{2.81580 26286 46091 20634 90}\textbf{123} E$-$02 [30]\\\underline{2.81580 26286 46091 20634 90048} E$-$02 [40] \end{tabular}\hspace{0.1cm}
\\
\hline
$2.0$ & $1$ & $0$ & $2.0$ & $1$ & $0$ & \begin{tabular}[c]{@{}l@{}}\underline{3.32260 90}\textbf{91}8 27546 98165 04057 E$-$02 [10]\\\underline{3.32260 90468 1}\textbf{83}02 55100 75237 E$-$02 [15]\\\underline{3.32260 90468 19031 39539 62493} E$-$02 [30]\\\underline{3.32260 90468 19031 39539 62493} E$-$02 [40]\\\underline{3.32260 9046}5 E$-$02{\footnotemark[1]} \end{tabular}\hspace{0.13cm}
\\
\hline
$2.1$ & $1$ & $0$ & $2.1$ & $1$ & $0$ & \begin{tabular}[c]{@{}l@{}}\underline{3.64394 52}\textbf{55}7 31327 16654 23236 E$-$02 [10]\\\underline{3.64394 52049 85}\textbf{24}8 40191 32912 E$-$02 [15]\\\underline{3.64394 52049 85324 78984 05}\textbf{22}7 E$-$02 [30]\\\underline{3.64394 52049 85324 78984 05166} E$-$02 [40] \end{tabular}\hspace{0.13cm}
\\
\hline
$2.2$ & $1$ & $0$ & $2.2$ & $1$ & $0$ & \begin{tabular}[c]{@{}l@{}}\underline{3.96639 94}\textbf{79}0 03807 01164 82993 E$-$02 [10]\\\underline{3.96639 94232 7}\textbf{11}20 86583 47853 E$-$02 [15]\\\underline{3.96639 94232 70337 92234 14}\textbf{51}1 E$-$02 [30]\\\underline{3.96639 94232 70337 92234 14397} E$-$02 [40] \end{tabular}\hspace{0.13cm}
\\
\hline
$2.3$ & $1$ & $0$ & $2.3$ & $1$ & $0$ & \begin{tabular}[c]{@{}l@{}}\underline{4.28635 9}\textbf{31}43 98419 21348 89833 E$-$02 [10]\\\underline{4.28635 92548 2}\textbf{97}50 51077 50463 E$-$02 [15]\\\underline{4.28635 92548 27904 76970 96}\textbf{42}0 E$-$02 [30]\\\underline{4.28635 92548 27904 76970 96264} E$-$02 [40]\end{tabular}
\footnotetext[1]{\cite{Rico1992}}
\footnotetext[2]{\cite{Peuker2008}}
\end{tabular}
\end{ruledtabular}
\end{table*}
\end{widetext}

\end{document}